\documentclass[10pt]{article}
\usepackage{amsmath}
\usepackage{bm}
\usepackage{lscape}
\usepackage{latexsym}
\usepackage[dvipdf]{graphicx}
\begin{document}
\title{Robust tests for ARCH in the presence of a misspecified conditional mean: 
a comparison of nonparametric approaches} 
\author{Daiki Maki\thanks{This research was supported by KAKENHI (Grant number: 16K03604). Address: Faculty of Commerce, Doshisha University, 
         Karasuma-higashi-iru, Imadegawa-dori, Kamigyo-ku, Kyoto Japan 602-8580 
         (E-mail:dmaki@mail.doshisha.ac.jp)}  \\ 
         Doshisha University
         \and Yasushi Ota \thanks{This research was supported by KAKENHI (Grant number: 18K03439). Faculty of Management, Okayama University of Science, 1-1 Ridaicyou, Okayama City, Okayama Japan 700-0005 
(E-mail:yota@mgt.ous.ac.jp)} \\ Okayama University of Science}
\date{}
\maketitle

\begin{center}
\large{Abstract}
\end{center}
\fontsize{11pt}{25pt}\selectfont
This study compares the statistical properties of autoregressive conditional heteroskedasticity (ARCH) tests that are robust to the presence of a misspecified conditional mean. 
The approaches employed are based on two nonparametric regressions for the conditional mean: 
an ARCH test with a Nadaraya-Watson kernel regression and  an ARCH test using a polynomial approximation regression. 
The two approaches do not require the specification of a conditional mean and can adapt to various nonlinear models, which are unknown a priori. 
The results reveal that the ARCH tests are robust to the misspecfied conditional mean models. 
The simulation results show that the ARCH tests based on the polynomial approximation regression approach have better statistical properties 
than those using the Nadaraya-Watson kernel regression approach for various nonlinear models. 
\\

\newpage
\fontsize{11pt}{27pt}\selectfont
\section{Introduction}
The presence of heteroskedasticity significantly impacts estimations and inferences in a time series analysis. 
Becker and Hurn (2009) and Pavlidis, Paya, and Peel (2010), for example, demonstrate that 
the presence of heteroskedasticity frequently leads to over-rejections of the null hypothesis  
when testing the null for the linearity of a conditional mean model against the alternative hypothesis of nonlinear time series models. 
Pavlidis, Paya, and Peel (2013) show that causality tests on the conditional mean demonstrate spurious causality relationships in the presence of multivariate heteroskedasticity. 
These facts indicate that tests for heteroskedasticity in data-generating processes (DGP) play an important role in time series analyses.

The most representative model for heteroskedasticity is Engle's (1982) autoregressive conditional heteroskedasticity (ARCH) model.  
ARCH is a simple and popular volatility model and continues to be widely used in the literature. 
When testing for heteroskedasticity, 
a regression model for the assumed conditional mean is first estimated. 
Next, ARCH is examined to use statistics such as the Lagrange multiplier (LM). 
If the conditional mean regression model is correctly specified, 
the ARCH test performs well. 
However, a misspecified conditional mean severely impedes the ARCH tests. 
Lumsdaine and Ng (1999) examine the properties of ARCH tests under a misspecified conditional mean. 
They show that the misspecification of the conditional mean over-rejects the null hypothesis for homoskedasticity. 
Similarly, Balke and Kapetanios (2007) clarify the influence of the neglected nonlinearity of the conditional mean on ARCH tests. 
Their analysis evidences the over-rejection of no ARCH effects when the nonlinearity of the conditional mean regression model is neglected. 
To appropriately test for ARCH, it is necessary to avoid the misspecified model of the conditional mean.

This study compares the statistical properties of ARCH tests that do not depend on the conditional mean model. 
The tests are applicable to various nonlinear conditional mean models and are robust to the misspecified conditional mean model. 
We employ two nonparametric approaches to avoid the misspecification of the conditional mean model. 
First is a regression using the Nadaraya-Watson kernel estimator, which is a representative nonparametric method. 
Nadaraya (1964) and Watson (1964) propose the method using a kernel density function in a regression analysis that does not depend on the model. 
McMillan (2001) and Exterkate, Groenen, Heij, and van Dijk (2016) show that the Nadaraya-Watson estimator is useful under various nonlinear models. 
Second is the regression analysis using a polynomial approximation of a general unknown nonlinear model. 
Stone (1977) and Katkovnik (1979) propose the local polynomial estimator on the basis of a polynomial approximation. 
Balke and Kapetanios (2007) develop a method to approximate unknown models using a neural network. 
P\'{e}guin-Feissolle, Strikholm, and Ter\"{a}svirta (2013) introduce a causality test that 
is based on a Taylor approximation of a general nonlinear model and is applicable to various nonlinear models.  
These approaches are relevant from the viewpoint of a polynominal approximation.

This study introduces ARCH tests using these nonparametric regression approaches to avoid the misspecification of the conditional mean 
and investigates the statistical properties of the introduced tests in various linear and nonlinear models. 
Erroneous ARCH tests based on misspecified conditional mean models and the failure to obtain sufficient reliability for the derived results increasingly impedes  
model constructions and statistical evaluation. 
Thus, it is important to clarify the influence of the introdued tests that do not depend on the model specification for various models.

In this study, we examine rejection frequencies under the null and alternative hypotheses for the introduced ARCH tests using Monte Carlo simulations. 
The simulation analyzes the influence of the lag length, the bandwidth selection for the Nadaraya-Watson estimator, and the approximation order for the polynominal approximation method on the results. 
The conditional mean models ivestigated in this study are linear autoregressive, threshold autoregressive, smooth transition autoregressive, Markov switching, and bi-linear models. 
These are popular nonlinear models used for empirical analysis and tend to cause spurious ARCH effects 
because it is difficult to distinguish between nonlinear models with homoskedastic variance and linear models with an ARCH effect. 
The Monte Carlo simulation results evidence that ARCH tests that are based on the polynomial approximation regression approach have better statistical properties 
than those using the Nadaraya-Watson kernel regression approach when DGP are various nonlinear models.

The remainder of this paper is organized as follows: 
Section 2 presents the influence of a misspecified conditional mean on the ARCH tests and proposes ARCH tests using nonparametric regression approaches for the conditional mean.  
Section 3 presents the statistical properties of the tests under nonlinear models. 
Section 4 concludes the paper.

\section{ARCH tests using nonparametric regression approaches for conditional mean}
We consider the following DGP with lag order $m$:  
\begin{equation}
             y_t=f(y_{t-1},\cdots, y_{t-m};\mbox{\boldmath $\beta$})+ u_{t}, \ \ \ t=1,\cdots,T
\end{equation}
where $f(\cdot;\cdot)$ is an unknown function and {\boldmath$\beta$} is a parameter vector. 
$u_t$ is a disturbance term with mean zero and variance denoted by
\begin{equation}
u_t=\sigma_t \epsilon_t; \ \ \ \sigma_t^2=\gamma_0+\sum_{i=1}^p \gamma_i u_{t-i}^2,  
\end{equation} 
where $\epsilon_t$ are independently and identically distributed (iid) random variables with mean zero and variance equal to one. 
Although the conditional variance could have model misspecification similar to the conditional mean, 
standard heteroskedastic tests have the ability to find linear ARCH effects even if the true conditional variance is generalized ARCH (GARCH) with or  without nonlinear parts. 
On the other hand, spurious ARCH effects tend to be observed when the conditional mean has model misspecifications. 
The misspecification of the conditional mean has clear impacts on the inference of variance, as shown by Lumsdaine and Ng (1999) and Balke and Kapetanios (2007). 
Thus, we focus on investigating the influence that the model misspecification of the conditional mean has on ARCH effects. 

The null hypothesis of homoskedasticity to test for the ARCH effect is denoted by 
\begin{equation}
H_0: \gamma_1=\cdots=\gamma_p=0,
\end{equation}
and the alternative hypothesis is 
\begin{equation}
H_1: \text{at least one} \ \gamma_i=0, \ \ \ i=1,\cdots,p.
\end{equation}
Even if we assume a GARCH model to be heteroskedastic, 
the testing procedure is the same as that in by Lee(1991) and Gel and Chen(2012). 
Therefore, we focus only on the ARCH test.  
Engle's (1982) standard ARCH test uses the auxiliary regression of squared residuals: 
\begin{equation}
\hat{u}_t^2=\gamma_0+\gamma_1 \hat{u}_{t-1}^2+\cdots+\gamma_p \hat{u}_{t-p}^2+\eta_t, 
\end{equation}
where $\eta_t$ is an error term. 
The LM test statistics is given by
\begin{equation}
LM=\frac{T\hat{d}^{\prime}\hat{W}(\hat{W}^{\prime}\hat{W})^{-1}\hat{W}^{\prime}\hat{d}}{\hat{d}^{\prime}\hat{d}},
\end{equation}
where $\hat{d}^{\prime}=(\hat{d}_1,\cdots,\hat{d}_T)$, $\hat{d}_t=(\hat{u}_t^2/\hat{\sigma}_u-1)$, $\hat{\sigma}^2=(1/T)\sum_{t=1}^T\hat{u}_t^2$, 
$\hat{W}^{\prime}=(\hat{w}_1,\cdots,\hat{w}_T)$, and $\hat{w}_t=(1,\hat{u}_{t-1}^2,\cdots,\hat{u}_{t-p}^2)$. 
The LM test statistic (6) is equivalent to $TR^2$, where $R^2$ is the coefficient for the determination of (5)$^1$ . 
Under the null hypothesis of no ARCH effects, the asymptotic distribution of (6) is $\chi^2(p)$.

When true DGP are denoted by (1), 
suppose that we estimate the following misspecified model: 
\begin{equation}
             y_t=g(y_{t-1},\cdots, y_{t-\tilde{m}};\mbox{\boldmath $\alpha$})+ u_{t},  
\end{equation}
where $g(\cdot;\cdot)$ is a misspecified function, $\tilde{m}$ is the lag length, and {\boldmath$\alpha$} is a parameter vector for the misspecified model. 
Accordingly, the residual is denoted by 
\begin{equation}
\hat{u}_t=u_t+f(y_{t-1},\cdots, y_{t-m};\mbox{\boldmath $\beta$})-\hat{g}(y_{t-1},\cdots, y_{t-m};\mbox{\boldmath $\alpha$})=u_t+e_t, 
\end{equation}
where $e_t=f(y_{t-1},\cdots, y_{t-m};\mbox{\boldmath $\beta$})-\hat{g}(y_{t-1},\cdots, y_{t-m};\mbox{\boldmath $\alpha$})$.  
The squared residual for $\hat{u}_t$ is 
\begin{equation}
\hat{u}_t^2=u_t^2+2u_t e_t+e^2_t. 
\end{equation}
Equation (9) means that the ARCH test correctly performs when $e_t \xrightarrow{p} 0$, 
whereas the ARCH test is subject to a model misspecification and leads to unreliable results when $e_t \xrightarrow{p} 0$ does not hold. 
For example, when true DGP (1) are a threshold autoregressive (TAR) model and misspecified estimation model (7) is a linear AR model, 
$e_t$ includes nonlinearity.  
As highlighted by Lumsdaine and Ng (1999) and Blake and Kapetanios (2007),  
such a misspecification results in a spurious ARCH effect. 
Therefore, a regression approach that does not depend on a specific model is necessary to avoid model misspecification and spurious ARCH effects.

The first approach that is robust to model misspecification is a nonparametric regression that is based on the Nadaraya-Watson kernel estimator. 
We consider the following conditional mean regression regression model: 
\begin{equation}
             y_t=m(y_{t-1},\cdots, y_{t-s})+ u_{t}, \ \ \ t=1,\cdots,T,
\end{equation}
where $m(\cdot)$ is the unknown regression function without any parametric form. 
The regression function for $y_t$ on $Y_t=(y_{t-1},\cdots, y_{t-s})^{\prime}$ is 
\begin{equation}
z(y_{t-1},\cdots,y_{t-s})=E(y_t|Y_t=y). 
\end{equation}
The most representative method to estimate the function is the Nadaraya-Watson estimator. 
The estimator is denoted by
\begin{equation}
\hat{z}(y_{t-1},\cdots, y_{t-s})=\frac{\sum_{t=1}^T K(\frac{Y_t-y}{h})y_t }{\sum_{t=1}^T K(\frac{Y_t-y}{h})},
\end{equation}
where $K(\frac{Y_t-y}{h})=K(\frac{y_{t-1}-y_1}{h_1})K(\frac{y_{t-2}-y_2}{h_2}) \cdots K(\frac{y_{t-s}-y_s}{h_{s}})$ is a product kernel function 
and $h$ denotes the bandwidth to determine the smoothness of the kernel function. 
Each kernel funcion satisfies the following: 
\begin{equation}
\int K(y)dy=1, \ \ \ \int yK(y)dy=0, \ \ \ \int y^2 K(y)dy>0. 
\end{equation}
This study uses the Gaussian kernel denoted by$^2$:  
\begin{equation}
K(\cdot)=\frac{1}{\sqrt{2\pi}}\exp(-\frac{y^2}{2}). 
\end{equation}

We use two bandwidth selections for $h$ that are derived by minimizing the integrated mean squared error (IMSE). 
First is Silverman's (1986) the plug-in method.  
The bandwidth obtained using the plug-in method is based on the following equation: 
\begin{equation}
h=c_0 T^{-1/1+s},
\end{equation}
where $c_0$ is a constant that depends on the kernel function and $s$ is the number of the regressor.  
When we use the Gaussian kernel, the optimal bandwidth selection is denoted by 
\begin{equation}
h_{opt} \approx 1.06\sigma T^{-1/(s+4)}, 
\end{equation}
where $\sigma$ is the standard deviation for $y_t$. 
The modified $h_{opt}$ that is robust to outliers is written as 
\begin{equation}
h=1.06 \min(\hat{\sigma}, \hat{Q}/1.34)T^{-1/(s+4)}, 
\end{equation}
where $\hat{Q}$ is the estimate for the interquartile range of $y_t$$^{3}$. 

Second is the cross-validation procedure developed by Rudemo (1982). 
When using the Gaussian kernel, we consider the following mean squeared error called the cross-validation criterion: 
\begin{equation}
CV(h)=\frac{1}{T}\sum_{i=1}^T (y_i-\hat{z}(Y_{-i}))^2,
\end{equation}
where $\hat{z}(Y_{-i})$ is a leave-one-out estimator that excludes $i$th observation.  
The optimal bandwidth $h$ for the cross-validation procedure is determined by minimizing $CV(h)$. 
Stone (1984) shows that bandwidth $h$ for the cross-validation can asymptotically select the optimal bandwidth 
from an IMSE viewpoint and has probability convergence to the bandwidth for the plug-in method. 
While bandwidth $h$ for the plug-in method depends on the assumed kernel density function, 
the cross-validation is not required to assume the kernel density function and 
can obtain a consistent estimator for the bandwidth that minimizes IMSE. 
It is possible that the residuals obtained using Nadaraya-Watson estimator (12) with bandwidth selection (17) or (18) have similar properties. 
Accordingly, the above-mentioned nonparametric regression approach is robust to the model misspecification of the conditional mean 
and thus, the ARCH test is correctly performed$^4$. 

The next approach adopted to avoid misspecification is a polynomial approximation of a general unknown nonlinear model. 
When we apply a $k$th-order Taylor approximation to true model (1), the regression model is denoted by  
\begin{equation}
y_t=\beta_0+\sum_{j=1}^q \beta_j y_{t-j}+\sum_{j_1=1}^q \sum_{j_2=j_1}^q \beta_{j_1j_2} \beta_{j_1 j_2} y_{t-j_1}y_{t-j_2}+\cdots+\sum_{j_1=1}^q \sum_{j_2=j_1}^q \cdots \sum_{j_k=j_{k-1}}^q \beta_{j_1\cdots j_k} y_{t-j_1}\cdots y_{t-j_k}+\epsilon_t,
\end{equation}
where $q$ is the lag length and $\epsilon_t$ is an error term that includes the remainder term of the Taylor series approximation. 
We assume $q \le k$ as a simple notation. 
If the true model is a linear AR model, all $\beta_{j_1 j_2}$ and $\beta_{j_1 \cdots j_k}$ are zero. 
In contrast, if the true model is nonlinear, one $\beta_{j_1 j_2}$ or $\beta_{j_1 \cdots j_k}$ is not zero at least. 
We investigate this using a standard Wald test. 
For example, (19) with $p=2$ and $k=2$ can be written as
\begin{equation}
y_t=\beta_0+\sum_{j=1}^2 \beta_j y_{t-j}+ \sum_{j_1=1}^2 \sum_{j_2=j_1}^2 b_{j_1j_2}y_{t-j_1}y_{t-j_2}+\epsilon_t. 
\end{equation}
The difference between the true model and the polynomial approximation regression model reduces because 
the polynomial regression can approximate various nonlinear models including the TAR and Markov switching models. 
When testing for ARCH effects under an unknown (true) model, 
ussing residuals obtained from polynomial approximation regression (19) can be advantageous 
since they show statistical properties similar to those of the true model. 
Therefore, the ARCH test using the residuals from the polynomial approximation regression does not appear to be influenced by model misspecification.

\section{Statistical properties of ARCH tests using nonparametric regression models}
This section examines the statistical properties of the ARCH tests using nonparametric regression models for the conditional mean presented in Section 2. 
We conduct Monte Carlo simulations to compare the rejection frequencies of the test statistics under various conditional mean models with and without ARCH effects. 
The simulations are based on 10,000 replications; a significance level of 5\%; and sample sizes with $T=100$, $250$, and $500$. 
To avoid the effect of initial conditions, data with $T+100$ are generated. 
We discard the initial 100 samples and use the data with sample size $T$. 
We compare ARCH tests (6) using the following regression models for the conditional mean: 
the AR model denoted $AR(p)$, polynomial approximation model (19) with second- and third-order Taylor approximation 
denoted as $T2(p)$ and $T3(p)$, 
and nonparametric regression model (12) with plug-in method (17) and cross validation method (18) denoted as $NP_{pl}(p)$ and $NP_{cv}(p)$. 
We set lag length $p$ to $p=1$ or $p=2^5$. 
The AR model is used as a benchmark for comparison.

First, we consider the following AR processes to examine the influence of lag length on the tests' performance. 
\begin{gather}
             y_t=\beta_0+ \beta_1 y_{t-1} +\beta_2 y_{t-2} + u_{t}, \\
             u_t =\sigma_t \epsilon_t, \\
             \sigma_t^2=\gamma_0+\gamma_1 u_{t-1}^2,
\end{gather}
where  $u_{t} \sim \text{i.i.d.}N(0,1)$. 
$\beta_0$ is set to $\beta_0=0$. 
Table 1 presents the rejection frequencies for the ARCH tests obtained from each regression model for the conditional mean. 
We use the following DGP: \\ 
DGP1-1: $y_t= 0.2 y_{t-1} + u_{t}$, \\
DGP1-2: $y_t= 0.7 y_{t-1} + u_{t}$, \\
DGP1-3: $y_t= 0.7 y_{t-1} -0.2 y_{t-2}+ u_{t}$, \\
DGP1-4: $y_t= 0.7 y_{t-1} -0.5 y_{t-2}+ u_{t}$. \\
These DGP have homoskedastic errors with $\gamma_0=1$ and $\gamma_1=0$ for (23). 
The rejection frequencies presented in Table 1 indicate the empirical size of the ARCH tests on the basis of each regression model.

For DGP1-1 and DGP1-2, which have lag order one, most of the tests have a small under-rejection but reasonable size performance, 
except for $NP_{pl}(2)$ and $NP_{cv}(2)$. 
$NP_{pl}(2)$ and $NP_{cv}(2)$ report over-rejections for DGP1-1 and DGP1-2. 
The rejection frequencies of $NP_{pl}(2)$ for DGP1-1 with $T=500$ and of $NP_{cv}(2)$ for DGP1-2 with $T=500$ are 0.143 and 0.101. 
An additional lag for the nonparametric regression of the conditional mean using  the Nadaraya-Watson estimator leads to size distortions in the ARCH tests. 
In contrast, $AR(2)$, $T2(2)$, and $T3(2)$ do not report over-rejections for DGP1-1 and DGP1-2. 
The results show that the additional lag for $AR$ and polynomial approximation regression do not impact the size of the ARCH tests. 
However, a lower lag length clearly influences the empirical size of all the tests. 
We see that the ARCH tests based on $AR(1)$, $T2(1)$, $T3(1)$, $NP_{pl}(1)$, and $NP_{cv}(1)$ over-reject the null hypothesis of homoskedastic variance 
under DGP1-3 or DGP1-4, which have a lag order of two. 
For example, the rejection frequencies of $AR(1)$, $T2(1)$, $T3(1)$, $NP_{pl}(1)$, and $NP_{cv}(1)$ for DGP1-4 with $T=250$ are 
0.127, 0.116, 0.097, 0.115, and 0.113, respectively. 
The size distortions in DGP1-4 are greater than those in DGP1-3.  
The influence of the lower lag length on the empirical size depends on the persistence parameter of DGP. 
Compared with the size distortions for the model with a lower lag length, 
those for the model with an additional lag length are smaller. 
Accordingly, we present the statistical properties for the models with two lags.

We examine the empirical size of the ARCH tests under the following conditional mean generated by the TAR models.\\ 
DGP2-1: $y_t=(0.7 y_{t-1}- 0.2 y_{t-2})I(y_{t-1}\ge0)+(0.1 y_{t-1}- 0.2 y_{t-2})I(y_{t-1} <0) + u_{t}$, \\
DGP2-2: $y_t=(0.7 y_{t-1}- 0.2 y_{t-2})I(y_{t-1}\ge0)+(-0.5 y_{t-1}- 0.2 y_{t-2})I(y_{t-1} <0) + u_{t}$ , \\
DGP2-3: $y_t=(0.7 y_{t-1}+ 0.2 y_{t-2})I(y_{t-1}\ge0)+(0.7 y_{t-1}- 0.7 y_{t-2})I(y_{t-1} <0) + u_{t}$, \\
DGP2-4: $y_t=(0.7 y_{t-1}- 0.2 y_{t-2})I(\Delta y_{t-1}\ge0)+(0.1 y_{t-1}- 0.2 y_{t-2})I(\Delta y_{t-1} <0) + u_{t}$, \\
DGP2-5: $y_t=(0.7 y_{t-1}- 0.2 y_{t-2})I(\Delta y_{t-1}\ge0)+(-0.5 y_{t-1}- 0.2 y_{t-2})I(\Delta y_{t-1} <0) + u_{t}$, \\
DGP2-6: $y_t=(0.7 y_{t-1}+ 0.2 y_{t-2})I(\Delta y_{t-1}\ge0)+(0.7 y_{t-1}- 0.7 y_{t-2})I(\Delta y_{t-1} <0) + u_{t}$, \\
where $I(\cdot)$ is an indicator function that takes the value of 1 if $I(\cdot)$ is ture and 0 if $I(\cdot)$ is not true. 
$u_t$ denotes a homoskedastic error similar to that from DGP1-1 to 1-4. 
While DGP2-1, 2-2, and 2-3 are standard TAR models whose indicator functions depend on $y_{t-1}$, 
DGP2-4, 2-5, and 2-6 are momentum threshold autoregressive (MTAR) models wherein the threshold is the difference $\Delta y_{t-1}$. 
These TAR models allow for asymmetric adjustments. 
In addtion, MTAR can capture the spiky properties of the process.

Figures 1 and 2 illustrate the sample path for DGP2-1 with homoskedastic errors and the ARCH effect $\gamma_0=0.3$ for (23). 
Figure 2 clearly shows the volatile behavior generated by the ARCH effect. 
However, Figure 3 illustrates that the sample path for DGP2-3 demonstrates a similar volatile movement even if the error is homoskedastic. 
As shown in figures 2 and 3, it is generally difficult to distinguish between the nonlinear conditional mean model with the homoskedastic error and the linear AR model with ARCH effect.  
Such a similarlity between the TAR model with homoskedastic errors and the linear AR model with ARCH effects may produce spurious statistical properties. 

Table 2 tabulates the simulation results. 
$AR(2)$ reports over-rejections for the null hypothesis of no ARCH effects. 
For DGP2-2 and DGP2-5, which have strong asymmetry, the size distortions of $AR(2)$ are significantly large. 
These results indicate that the use of the AR model for the conditional mean leads to spurious ARCH effects 
when the true DGP are based on the TAR or MTAR model. 
In additon, the over-rejections increase with a large sample size. 
Unlike the performance of $AR(2)$, 
the polynomial approximation regression models $T2(2)$ and $T3(2)$ and nonparametric regression models $NP_{pl}(2)$ and $NP_{cv}(2)$ perform better. 
For example, the rejection frequencies of $AR(2)$, $T2(2)$, $T3(2)$, $NP_{pl}(2)$, and $NP_{cv}(2)$ for DGP2-2 with $T=250$ are 
0.373, 0.040, 0.033, 0.042, and 0.051, respectively. 
For $T3(2)$, on the other hand, the rejection frequency is 0.058. 
$T3(2)$ has a more reasonable size compared with those for $T2(2)$, $NP_{cl}(2)$, and $NP_{cv}(2)$. 
$T2(2)$, $NP_{cl}(2)$, and $NP_{cv}(2)$ report size distortions in certain cases. 
The rejection frequencies of $T2(2)$, $NP_{cl}(2)$, and $NP_{cv}(2)$ for DGP2-3 with $T=500$ are 0.096, 0.139, and 0.104, respectively. 
Thus, the polynomial approximation regression model $T3(2)$ is a more appropriate approach to test for ARCH than other approaches under the TAR or MTAR model.

Table 3 presents the rejection frequencies for each test under smooth transition autoregressive (STAR) models generated by the followings:\\ 
DGP3-1: $y_t=0.7 y_{t-1}- 0.2 y_{t-2}+(-0.5 y_{t-1}- 0.2 y_{t-2})(1-\exp (-0.1y_{t-1}^2)) + u_{t}$, \\
DGP3-2: $y_t=0.7 y_{t-1}- 0.2 y_{t-2}+(-y_{t-1}- 0.2 y_{t-2})(1-\exp (-0.1y_{t-1}^2)) + u_{t}$, \\
DGP3-3: $y_t=0.7 y_{t-1}- 0.2 y_{t-2}+(-y_{t-1}- 0.2 y_{t-2})(1-\exp (-y_{t-1}^2))  + u_{t}$, \\
DGP3-4: $y_t=0.7 y_{t-1}- 0.2 y_{t-2}+(-0.5 y_{t-1}- 0.2 y_{t-2})(1+\exp (-0.1y_{t-1}))^{-1} + u_{t}$, \\
DGP3-5: $y_t=0.7 y_{t-1}- 0.2 y_{t-2}+(- y_{t-1}- 0.2 y_{t-2})(1+\exp (-0.1y_{t-1}))^{-1} + u_{t}$, \\
DGP3-6: $y_t=0.7 y_{t-1}- 0.2 y_{t-2}+(- y_{t-1}- 0.2 y_{t-2})(1+\exp (-y_{t-1}))^{-1} + u_{t}$, \\
where $u_t$ denotes homoskedastic errors similar to those in tables 1 and 2. 
STAR models have the time-varying properties of the conditional mean. 
DGP3-1, 3-2, and 3-3 impose symmetry constraints on the time-varying properties, 
whereas DGP3-4, 3-5, and 3-6, which are logistic STAR models, allow asymmetry. 
DGP3-2 and 3-5 produce a smoother and more marginal change than DGP3-3 and 3-6. 
We observe that $AR(2)$, $T2(2)$, and $NP_{pl}(2)$ partially reject the null hypothesis of no ARCH effects.   
The rejection frequencies of $AR(2)$ is higher than those of the other regression models for DGP3-2 and 3-6.  
$T2(2)$ shows size distortions for DGP3-2.  
$NP_{pl}(2)$ reports a slight over-rejection with $T=500.$ 
In contrast, the shape of the transition function does not have a clear impact on the empirical size of $T3(2)$ and $NP_{cv}(2)$. 
$T3(2)$ and $NP_{cv}(2)$ can capture the properties of STAR models and allows the ARCH test to perform well.

In addition, we present the results of each test for the other nonlinear processes: \\
DGP4-1: $y_t=(0.7 y_{t-1}- 0.2 y_{t-2})s_t+(0.3 y_{t-1}- 0.2 y_{t-2})(1-s_t) + u_{t}$, \ \ $p_{00}=p_{11}=0.7$, \\
DGP4-2: $y_t=(0.7 y_{t-1}- 0.2 y_{t-2})s_t+(0.3 y_{t-1}- 0.2 y_{t-2})(1-s_t) + u_{t}$ , \ \ $p_{00}=p_{11}=0.98$, \\
DGP4-3: $y_t=(0.7 y_{t-1}+ 0.2 y_{t-2})s_t+(0.3 y_{t-1}- 0.2 y_{t-2})(1-s_t)  + u_{t}$, \ \ $p_{00}=p_{11}=0.98$,  \\
DGP4-4: $y_t=0.1y_{t-1}u_{t-1} +0.1 y_{t-2}u_{t-2} + u_{t}$, \\
DGP4-5: $y_t=0.3y_{t-1}u_{t-1} +0.1 y_{t-2}u_{t-2} + u_{t}$, \\
DGP4-6: $y_t=0.1y_{t-1}u_{t-1} -0.1 y_{t-2}u_{t-2} + u_{t}$, \\
where $u_{t} \sim \text{i.i.d.}N(0,1)$ and $s_t$ is a random variable that takes the value of 0 or 1. 
DGP4-1, 4-2, and 4-3 are Markov switching processes and $s_t$ determines the behavior. 
Whether $s_t$ takes the value of 0 or 1 depends on the transition probabilities $p_{11}$ and $p_{00}$. 
$p_{11}=P(s_{t+1}=1|s_t=1)$ denotes the change probability from the state $s_t=1$ to state $s_{t+1}=1$. 
Similarly, the transition probabilities are denoted by $p_{00}=P(s_{t+1}=0|s_t=0)$, $p_{10}=1-p_{00}=P(s_{t+1}=1|s_t=0)$, and $p_{01}=1-p_{11}=P(s_{t+1}=0|s_t=1)$, respectively. 
They are set to $p_{11}=p_{00}=0.7$ for DGP4-1 and $p_{11}=p_{00}=0.98$ for DGP4-2 and 4-3. 
While DGP4-1 has frequent switches in the AR parameters, DGP4-2 and 4-3 show persistent switches.  
DGP4-4, 4-5, and 4-6 are bilinear models that are used to model rare, volatile, or outburst processes.

$AR(2)$ that neglects nonlinearity causes spurious ARCH effect, which are similar to the results in tables 2 and 3. 
The results for the nonparametric regression models using the Nadaraya-Watson estimator depend on the bandwidth selection. 
$NP_{pl}(2)$ under-rejects the null hypothesis for DGP4-1, DGP4-2, and DGP4-5 and over-rejects that for DGP4-3, 4-4, and 4-6. 
$NP_{cv}(2)$ performs well for DGP4-2, DGP4-4, and DGP4-6 and over-rejects the null hypothesis for DGP4-1, 4-3, and 4-5. 
$T2(2)$ has relatively reasonable emirical sizes for $T=100$ and $200$, but reports size distortions for DGP4-1, 4-2, 4-3, 4-5, and 4-6 with $T=500$. 
Here as well, we find that $T3(2)$ generally performs better.

The simulation results from tables 1 to 4 evidence that the model misspecification of the conditional mean causes size distiortions for the null hypothesis of no ARCH effects. 
The ARCH tests using the AR regression model are sensitive to the presence of the nonlinear conditional mean and show high over-rejections. 
This can be attributed by neglected nonlinearity and difficulties in distinguishing between the nonlinearity of the conditional mean and the ARCH effects. 
While the noparametric regression models using the Nadaraya-Watson estimator partially perform well,  
the rejection frequencies strongly depend on DGP and the bandwidth selection. 
By contrast, the size properties of $T3(2)$ outperform those of other models and are close to the nominal size at 5\%. 
Therefore,  $T3(2)$ can approximate the (unknown) linear and nonlinear conditional mean models well and produce reliable ARCH tests.

Tables 5 and 6 report the nominal power and size-corrected power properties for the ARCH tests. 
We use DGP1-3, DGP2-1, 2-4, 3-1, 3-4, 4-1, and 4-4 for power comparison. 
Each DGP has an ARCH effect denoted by
\begin{gather}
             u_t =\sigma_t \epsilon_t, \\
             \sigma_t^2=\gamma_0+\gamma_1 u_{t-1}^2, 
\end{gather}
where $\gamma_0$ and $\gamma_1$ are set to $\gamma_0=1$ and $\gamma_1=(0.1,0.3)$.   
The powers of $AR(2)$ are clearly higher than those of other models in Table 5. 
We have a relatively reasonable evaluation of the power for DGP1-3 
because the size properties of $AR(2)$ and other tests are close to the nominal level 0.05 (Table 1). 
However, we cannot correctly evaluate the high nominal powers of $AR(2)$ for other DGP. 
The higher powers of $AR(2)$ are influenced by size distortions presented in tables from 2-4. 
The power properties of the nonparametric models are more appropriately interpreted  
because $T2(2)$ and $T3(3)$ do not over-reject the null hypothesis for DGP in Table 5 and the size distortions of $NP_{pl}(2)$ and $NP_{cv}(2)$ are smaller than those of $AR(2)$. 
In comparison, we observe that the polynomial approximation models $T2(2)$ and $T3(2)$ perform better than $NP_{pl}(2)$ and $NP_{cv}(2)$.  
Note that the powers of $NP_{pl}$(2) are quite small when the ARCH effect is $\gamma_1=0.1$. 
For $\gamma_1=0.3$, the nonparametric regression models report sufficient power to identify the ARCH effects.

We compare the power properties among the models without the influences of size distortions. 
Table 6 demonstrates the size-corrected power. 
The powers of $AR(2)$ in Table 6 are lower than those in Table 5 because the size distortions are corrected. 
$AR(2)$ still performs well even if the size is corrected.  
The ability to detect ARCH effects in the nonlinear models for $T2(2)$ is high, similar to that of $AR(2)$. 
While the powers of $T3(2)$ is slightly smaller than those of $T2(2)$ because $T3(2)$ has additional regression parameters for the conditional mean, 
it has sufficient power to find the ARCH effect. 
The rejection frequencies of $NP_{pl}(2)$ and $NP_{cv}(2)$ for $\gamma_1=0.1$ are inferior to those of other models in Table 6. 
While they relatively perform well for $\gamma_1=0.3$ with $T=100$, 
other models have better power properties, particularly for $T=250$ and $500$.

The comparison of the ARCH tests using each regression model for the conditional mean indicates that 
the presence of the nonlinear conditional mean has influences of size and power properties on the ARCH tests.  
The AR regression models  have higher over-rejection of the null hypothesis of no ARCH effects for the nonlinear conditional mean models.  
The ARCH tests based on AR models for the nonlinear conditional mean are not effective from the viewpoints of size and power. 
This is because size-corrected tests are needed and the true model is generally unknown a priori. 
The nonparametric regression models using the Nadaraya-Watson estimator tend to have slight size distortions and low power. 
The polynomial approximation model $T2(2)$ shows slight over-rejection depending on the nonlinear conditional mean and sample size, 
although it has better power properties for the ARCH effect with the nonlinear conditional mean. 
$T3(2)$ has reasonable size and power properties and yeilds reliable results for the ARCH tests irrespective of the conditional mean models.

\section{Summary and conclusion}
This study compares the statistical properties of the ARCH tests that are robust to misspecified conditional mean models. 
ARCH tests are important for statistical modeling because the presence of ARCH affects the statistical inference of the conditional mean regression model and the analysis of volatility. 
However, it is difficult to determine the correct specified conditional mean model and possible to employ a misspecified conditonal mean model.   
This may lead to unreliable results. 
Therefore, it is neccesary to compare robust ARCH tests to various unknown conditional mean model and clarify their statistical properties. 
The approaches employed in this study are based on two nonparametric regressions: 
an ARCH test using the Nadaraya-Watson kernel regression and  an ARCH test with the polynomial approximation. 
The two approches can adapt to various nonlinear models. 
Since a true model is generally unknown a priori, 
they are robust to misspecfied models. 
The Monte Carlo simulations evidence that the ARCH tests based on the polynomial regression approach have better statistical properties 
than those using the Nadaraya-Watson kernel regression approach for various nonlinear conditional mean models. 
In particular, the test using the regression approach based on the third-order Taylor approximation has a reasonable and acceptable size and sufficient power for any time series models.  
The results further show that the ARCH test using the polynomial approximation approach is useful 
when testing if DGP have an ARCH effect and for ARCH without model specifications when the conditional mean model is unknown a priori. 
Robust univariate and multivariate ARCH tests that do not depend on the model specification of the conditional variance in addition to the conditional mean are left for further study.

\newpage
\fontsize{10pt}{20pt}\selectfont
\begin{flushleft}
{\Large Footnotes}\\
\end{flushleft} 
1. Catani and Ahlgren (2017) propose an LM test for ARCH using high-dimentional vector autoregressive models. 
In addition, Gel and Chen (2012) introduce bootstrap ARCH tests. 
\\
\\
2. Other kernel functions include uniform, Epanechnikov, biweight, and triweight kernel functions. 
In general, while the type of kernel functions does not have a large impact on the estimation results, 
the selection of bandwidth significantly influences the estimation results. 
\\
\\
3. Sneather and Jones (1991) propose another bandwidth selection that is based on the plug-in method. 
\\ 
\\ 
4. Shimizu (2014) introduces the estimation of nonparametric AR(1)-ARCH(1) using wild bootstrap. 
Shin and Hwang (2015) apply stationary bootstrap to estimate nonparametric AR(1)-ARCH(1). 
\\ 
\\ 
5. Zambom and Kim (2017) propse lag selection in the nonparametric conditional heteroskedastic models. 
Compared to conventional methods, 
this method more appropriately selects lag length for various nonlinear models. 
We fix lag length in this paper to investigate the statistical performance of the nonparametric regression models.  
\\
\\

\newpage
\fontsize{11pt}{20pt}\selectfont

\newpage
\fontsize{11pt}{15pt}\selectfont
\begin{landscape}
\begin{center}
Table 1: Rejection frequencies under AR models 
\end{center}

\begin{center}
\begin{tabular}{c|cccccccccc} \hline       
                    &$AR(1)$&$AR(2)$& $T2(1)$& $T2(2)$ &$T3(1)$& $T3(2)$&$NP_{pl}(1)$&$NP_{pl}(2)$&$NP_{cv}(1)$&$NP_{cv}(2)$\\ \hline 
DGP1-1         &           &           &           &             &           &           &               &                &                &                 \\
$T=100$        & 0.035   & 0.035  & 0.035   &  0.025    &  0.032  & 0.028   & 0.031      & 0.041        &  0.040      &  0.034        \\
$T=250$        & 0.044   & 0.040  & 0.038   &  0.040   & 0.043    & 0.037   & 0.046      & 0.093        &  0.051      & 0.042         \\
$T=500$        & 0.041   & 0.043  & 0.046   &  0.041   & 0.046    & 0.046   & 0.051      & 0.143        &  0.052      & 0.047         \\  
DGP1-2         &           &           &           &            &            &            &              &                &                &                 \\
$T=100$        & 0.036   & 0.032  & 0.035   &  0.029    &  0.029  & 0.034   & 0.034      & 0.026        &  0.030      &  0.031        \\
$T=250$        & 0.039   & 0.040  & 0.038   &  0.038   & 0.041    & 0.043   & 0.037      & 0.025        &  0.039      & 0.066         \\
$T=500$        & 0.042   & 0.045  & 0.041   &  0.039   & 0.042    & 0.044   & 0.041      & 0.029        &  0.044      & 0.101         \\  
DGP1-3         &           &           &           &             &           &           &               &                &                &                 \\
$T=100$        & 0.042   & 0.031  & 0.032   &  0.027   &  0.032  & 0.031   & 0.035      & 0.018        &  0.042      &  0.030        \\
$T=250$        & 0.052   & 0.042  & 0.046   &  0.035   & 0.047    & 0.040   & 0.049      & 0.026        &  0.046      & 0.049         \\
$T=500$        & 0.063   & 0.044  & 0.061   &  0.040   & 0.054    & 0.045   & 0.064      & 0.037        &  0.052      & 0.075         \\  
DGP1-4         &           &           &           &            &            &            &              &                &                &                 \\
$T=100$        & 0.063   & 0.034  & 0.060   &  0.026   &  0.048  & 0.028   & 0.051      & 0.019        &  0.072      &  0.029        \\
$T=250$        & 0.127   & 0.043  & 0.116   &  0.037   & 0.097    & 0.038   & 0.115      & 0.027        &  0.113      & 0.048         \\
$T=500$        & 0.213   & 0.041  & 0.201   &  0.042   & 0.189    & 0.040   & 0.208      & 0.032        &  0.181      & 0.069         \\  
    \hline 
   \end{tabular}
\end{center}
\end{landscape}

\newpage
\begin{center}
Table 2: Rejection frequencies under TAR models 
\end{center}

\begin{center}
\begin{tabular}{c|ccccc} \hline          
                    &$AR(2)$&$T2(2)$& $T3(2)$&$NP_{pl}(2)$&$NP_{cv}(2)$ \\ \hline 
DGP2-1        &            &           &           &               &                 \\
$T=100$        & 0.049   & 0.029  & 0.029   &  0.028      &  0.030        \\
$T=250$        & 0.080   & 0.036  & 0.040   &  0.051      & 0.041         \\
$T=500$        & 0.116   & 0.046  & 0.041   &  0.081      & 0.071         \\  
DGP2-2        &            &           &           &               &                 \\
$T=100$        & 0.150   & 0.029  & 0.025   &  0.021      &  0.033        \\
$T=250$        & 0.373   & 0.040  & 0.033   &  0.042      & 0.051        \\
$T=500$        & 0.658   & 0.055  & 0.040   &  0.069      & 0.073         \\  
DGP2-3        &            &           &           &               &                 \\
$T=100$        & 0.083   & 0.038  & 0.026   &  0.104      &  0.026        \\
$T=250$        & 0.144   & 0.064  & 0.040   &  0.132      & 0.065         \\
$T=500$        & 0.224   & 0.096  & 0.058   &  0.139      & 0.104         \\  
DGP2-4        &            &           &           &               &                 \\
$T=100$        & 0.063   & 0.027  & 0.027   &  0.034      &  0.030        \\
$T=250$        & 0.117   & 0.040  & 0.038   &  0.086      & 0.046         \\
$T=500$        & 0.188   & 0.045  & 0.041   &  0.135      & 0.073         \\  
DGP2-5        &            &           &           &               &                 \\
$T=100$        & 0.359   & 0.039  & 0.026   &  0.027      &  0.053        \\
$T=250$        & 0.748   & 0.083  & 0.051   &  0.059      & 0.076         \\
$T=500$        & 0.962   & 0.150  & 0.090   &  0.094      & 0.144         \\  
DGP2-6        &            &           &           &               &                 \\
$T=100$        & 0.081   & 0.036  & 0.028   &  0.030      &  0.040        \\
$T=250$        & 0.176   & 0.050  & 0.044   &  0.037      & 0.075         \\
$T=500$        & 0.277   & 0.069  & 0.052   &  0.044      & 0.142         \\  
                    &           &          &            &               &                  \\
    \hline
   \end{tabular}
\end{center}

\newpage
\begin{center}
Table 3: Rejection frequencies under STAR models 
\end{center}
\begin{center}

\begin{tabular}{c|ccccc} \hline          
                    &$AR(2)$&$T2(2)$& $T3(2)$&$NP_{pl}(2)$&$NP_{cv}(2)$ \\ \hline 
DGP3-1        &            &           &           &               &                 \\
$T=100$        & 0.053   & 0.032  & 0.033   &  0.023      &  0.027        \\
$T=250$        & 0.068   & 0.045  & 0.038   &  0.053      & 0.043        \\
$T=500$        & 0.094   & 0.068  & 0.043   &  0.088      & 0.063         \\  
DGP3-2        &            &           &           &               &                 \\
$T=100$        & 0.088   & 0.043  & 0.029   &  0.028      &  0.028        \\
$T=250$        & 0.184   & 0.107  & 0.038   &  0.061      & 0.042       \\
$T=500$        & 0.303   & 0.238  & 0.044   &  0.088      & 0.054         \\  
DGP3-3        &            &           &           &               &                 \\
$T=100$        & 0.042   & 0.024  & 0.026   &  0.027      &  0.031        \\
$T=250$        & 0.048   & 0.038  & 0.030   &  0.054      & 0.047        \\
$T=500$        & 0.057   & 0.046  & 0.042   &  0.085      & 0.069         \\  
DGP3-4        &            &           &           &               &                 \\
$T=100$        & 0.035   & 0.030  & 0.031   &  0.027      &  0.031        \\
$T=250$        & 0.044   & 0.034  & 0.040   &  0.055      & 0.043         \\
$T=500$        & 0.045   & 0.040  & 0.043   &  0.082      & 0.049         \\  
DGP3-5        &            &           &           &               &                 \\
$T=100$        & 0.032   & 0.029  & 0.025   &  0.034      &  0.028        \\
$T=250$        & 0.042   & 0.039  & 0.040   &  0.067      & 0.043        \\
$T=500$        & 0.053   & 0.041  & 0.042   &  0.111      & 0.047         \\  
DGP3-6        &            &           &           &               &                 \\
$T=100$        & 0.109   & 0.029  & 0.029   &  0.028      &  0.032        \\
$T=250$        & 0.273   & 0.042  & 0.037   &  0.049      & 0.042         \\
$T=500$        & 0.483   & 0.048  & 0.042   &  0.073      & 0.055         \\  
                    &           &          &            &               &                  \\
    \hline
   \end{tabular}

\end{center}

\newpage
\begin{center}
Table 4: Rejection frequencies under MS and bilinear models 
\end{center}
\begin{center}

\begin{tabular}{c|ccccc} \hline          
                    &$AR(2)$&$T2(2)$& $T3(2)$&$NP_{pl}(2)$&$NP_{cv}(2)$ \\ \hline 
DGP4-1        &            &           &           &               &                 \\
$T=100$        & 0.051   & 0.034  & 0.029   &  0.021      &  0.049        \\
$T=250$        & 0.092   & 0.064  & 0.040   &  0.025      & 0.065       \\
$T=500$        & 0.137   & 0.101  & 0.068   &  0.029      & 0.077         \\  
DGP4-2        &            &           &           &               &                 \\
$T=100$        & 0.047   & 0.029  & 0.026   &  0.023      &  0.040        \\
$T=250$        & 0.076   & 0.053  & 0.037   &  0.026      & 0.051       \\
$T=500$        & 0.108   & 0.084  & 0.055   &  0.034      & 0.057         \\  
DGP4-3        &            &           &           &               &                 \\
$T=100$        & 0.048   & 0.035  & 0.031   &  0.112      &  0.051        \\
$T=250$        & 0.084   & 0.054  & 0.044   &  0.156      & 0.073        \\
$T=500$        & 0.116   & 0.087  & 0.057   &  0.141      & 0.100         \\  
DGP4-4        &            &           &           &               &                 \\
$T=100$        & 0.066   & 0.026  & 0.028   &  0.034      &  0.049        \\
$T=250$        & 0.112   & 0.036  & 0.038   &  0.076      & 0.051         \\
$T=500$        & 0.170   & 0.043  & 0.044   &  0.129      & 0.052         \\  
DGP4-5        &            &           &           &               &                 \\
$T=100$        & 0.456   & 0.035  & 0.032   &  0.027      &  0.107        \\
$T=250$        & 0.851   & 0.071  & 0.040   &  0.033      & 0.110        \\
$T=500$        & 0.987   & 0.100  & 0.046   &  0.036      & 0.133         \\  
DGP4-6        &            &           &           &               &                 \\
$T=100$        & 0.053   & 0.029  & 0.030   &  0.037      &  0.040        \\
$T=250$        & 0.090   & 0.040  & 0.039   &  0.078      & 0.051         \\
$T=500$        & 0.134   & 0.040  & 0.046   &  0.133      & 0.056         \\  
                    &           &          &            &               &                  \\
    \hline
   \end{tabular}

\end{center}

\begin{landscape}
\begin{center}
Table 5: Nominal power properties for ARCH tests 
\end{center}

\begin{center}
\begin{tabular}{c|ccccc|ccccc} \hline       
                    &\multicolumn{5}{|c|}{$\gamma_1=0.1$} & \multicolumn{5}{|c}{$\gamma_1=0.3$}   \\ \hline
                    &$AR(2)$&$T2(2)$& $T3(2)$&$NP_{pl}(2)$&$NP_{cv}(2)$&$AR(2)$&$T2(2)$&$T3(2)$&$NP_{pl}(2)$&$NP_{cv}(2)$\\ \hline 
DGP1-3         &           &           &           &             &           &           &               &              &             &                 \\
$T=100$        & 0.119   & 0.070  & 0.032   &  0.043    &  0.055  & 0.440   & 0.294      & 0.138        &  0.218      &  0.205        \\
$T=250$        & 0.267   & 0.197  & 0.122   &  0.082   & 0.089    & 0.848   & 0.777      & 0.646        &  0.589      & 0.493         \\
$T=500$        & 0.486   & 0.413  & 0.309   &  0.182   & 0.164    & 0.990   & 0.982      & 0.959        &  0.913      & 0.845         \\  
DGP2-1         &           &           &           &            &            &            &              &                &                &                 \\
$T=100$        & 0.146   & 0.065  & 0.032   &  0.027    &  0.065  & 0.444   & 0.290      & 0.143        &  0.159      &  0.243        \\
$T=250$        & 0.364   & 0.213  & 0.119   &  0.046   & 0.102    & 0.868   & 0.787      & 0.673        &  0.484      & 0.552         \\
$T=500$        & 0.624   & 0.429  & 0.325   &  0.108   & 0.192    & 0.991   & 0.980      & 0.963        &  0.877      & 0.879         \\  
DGP2-4         &           &           &           &             &           &           &               &                &                &                 \\
$T=100$        & 0.164   & 0.068  & 0.038   &  0.018   &  0.072  & 0.446   & 0.284      & 0.147        &  0.122      &  0.239        \\
$T=250$        & 0.400   & 0.218  & 0.127   &  0.029   & 0.099    & 0.853   & 0.774      & 0.642        &  0.408      & 0.516         \\
$T=500$        & 0.677   & 0.446  & 0.337   &  0.065   & 0.158    & 0.989   & 0.980      & 0.958        &  0.792      & 0.844         \\  
DGP3-1         &           &           &           &            &            &            &              &                &                &                 \\
$T=100$        & 0.164   & 0.081  & 0.029   &  0.027   &  0.057  & 0.474   & 0.315      & 0.150        &  0.142     &  0.245        \\
$T=250$        & 0.363   & 0.275  & 0.119   &  0.047   & 0.111   & 0.879   & 0.817      & 0.647        &  0.477      & 0.609         \\
$T=500$        & 0.617   & 0.540  & 0.320   &  0.105   & 0.224    & 0.992   & 0.986      & 0.964        &  0.858      & 0.917         \\  
DGP3-4         &           &           &           &            &           &           &               &                &                &                 \\
$T=100$        & 0.115   & 0.062  & 0.033   &  0.024   &  0.072  & 0.440   & 0.287      & 0.143        &  0.159      &  0.282        \\
$T=250$        & 0.275   & 0.203  & 0.116   &  0.051   & 0.141    & 0.850   & 0.777      & 0.655        &  0.486      & 0.647         \\
$T=500$        & 0.490   & 0.418  & 0.325   &  0.109   & 0.258    & 0.988   & 0.982      & 0.964        &  0.859      & 0.933         \\  
DGP4-1         &           &           &           &            &            &            &              &                &                &                 \\
$T=100$        & 0.154   & 0.083  & 0.038   &  0.034   &  0.113  & 0.456  & 0.297      & 0.143        &  0.178      &  0.323        \\
$T=250$        & 0.379   & 0.294  & 0.180   &  0.086   & 0.235    & 0.863   & 0.797      & 0.662        &  0.532      & 0.696         \\
$T=500$        & 0.654   & 0.586  & 0.466   &  0.213   & 0.431    & 0.990   & 0.985      & 0.973        &  0.891      & 0.948         \\    
DGP4-4         &           &           &           &            &            &            &              &                &                &                 \\
$T=100$        & 0.189   & 0.067  & 0.030   &  0.023   &  0.117  & 0.481   & 0.283      & 0.135        &  0.140      &  0.369        \\
$T=250$        & 0.425   & 0.201  & 0.111   &  0.036   & 0.215    & 0.884   & 0.781      & 0.639        &  0.430      & 0.715         \\
$T=500$        & 0.693   & 0.419  & 0.310   &  0.067   & 0.354    & 0.991   & 0.983      & 0.964        &  0.828      & 0.950          \\
                   &           &          &            &            &              &           &              &                &                &                 \\
    \hline 
   \end{tabular}
\end{center}
\end{landscape}

\begin{landscape}
\begin{center}
Table 6: Size-corrected power properties for ARCH tests 
\end{center}

\begin{center}
\begin{tabular}{c|ccccc|ccccc} \hline       
                    &\multicolumn{5}{|c|}{$\gamma_1=0.1$} & \multicolumn{5}{|c}{$\gamma_1=0.3$}   \\ \hline
                    &$AR(2)$&$T2(2)$& $T3(2)$&$NP_{pl}(2)$&$NP_{cv}(2)$&$AR(2)$&$T2(2)$&$T3(2)$&$NP_{pl}(2)$&$NP_{cv}(2)$\\ \hline 
DGP1-3         &           &           &           &             &           &           &               &              &             &                 \\
$T=100$        & 0.145   & 0.088  & 0.047   &  0.075    &  0.081  & 0.479  & 0.325      & 0.165        &  0.280      &  0.235        \\
$T=250$        & 0.285   & 0.235  & 0.127   &  0.114   & 0.092    & 0.862   & 0.808      & 0.665        &  0.644      & 0.494         \\
$T=500$        & 0.501   & 0.429  & 0.321   &  0.191   & 0.127    & 0.990   & 0.982      & 0.967        &  0.919      & 0.807         \\  
DGP2-1         &           &           &           &            &            &            &              &                &                &                 \\
$T=100$        & 0.162   & 0.094  & 0.048   &  0.041    &  0.080  & 0.460   & 0.343      & 0.177        &  0.203      &  0.261        \\
$T=250$        & 0.296   & 0.235  & 0.138   &  0.044   & 0.107    & 0.820   & 0.811      & 0.678        &  0.472      & 0.571         \\
$T=500$        & 0.476   & 0.453  & 0.342   &  0.078   & 0.168    & 0.980   & 0.984      & 0.968        &  0.819      & 0.864         \\  
DGP2-4         &           &           &           &             &           &           &               &                &                &                 \\
$T=100$        & 0.148   & 0.089  & 0.058   &  0.028   &  0.086  & 0.417   & 0.331      & 0.185        &  0.133      &  0.271        \\
$T=250$        & 0.279   & 0.240  & 0.149   &  0.020   & 0.088   & 0.763   & 0.786      & 0.669        &  0.348      & 0.500         \\
$T=500$        & 0.412   & 0.457  & 0.352   &  0.032   & 0.120    & 0.951   & 0.979      & 0.966        &  0.719      & 0.803         \\  
DGP3-1         &           &           &           &            &            &            &              &                &                &                 \\
$T=100$        & 0.165   & 0.107  & 0.046   &  0.041   &  0.098  & 0.470   & 0.366      & 0.179        &  0.188     &  0.293        \\
$T=250$        & 0.315   & 0.262  & 0.129   &  0.041   & 0.156   & 0.849   & 0.815      & 0.668        &  0.472      & 0.620         \\
$T=500$        & 0.502   & 0.483  & 0.342   &  0.075   & 0.246    & 0.984   & 0.982      & 0.971        &  0.820      & 0.908         \\  
DGP3-4         &           &           &           &            &           &           &               &                &                &                 \\
$T=100$        & 0.193   & 0.106  & 0.044   &  0.043   &  0.098  & 0.477   & 0.346      & 0.176        &  0.192      &  0.315        \\
$T=250$        & 0.387   & 0.295  & 0.132   &  0.046   & 0.150    & 0.865   & 0.808      & 0.674        &  0.475      & 0.674         \\
$T=500$        & 0.625   & 0.545  & 0.327   &  0.080   & 0.259    & 0.988   & 0.984      & 0.966        &  0.819      & 0.936         \\  
DGP4-1         &           &           &           &            &            &            &              &                &                &                 \\
$T=100$        & 0.153   & 0.115  & 0.059   &  0.063   &  0.117  & 0.448  & 0.345      & 0.182        &  0.236      &  0.326        \\
$T=250$        & 0.287   & 0.263  & 0.207   &  0.119   & 0.219    & 0.803   & 0.773      & 0.690        &  0.599      & 0.671         \\
$T=500$        & 0.451   & 0.452  & 0.415   &  0.267   & 0.391    & 0.972   & 0.967      & 0.960        &  0.919      & 0.937         \\    
DGP4-4         &           &           &           &            &            &            &              &                &                &                 \\
$T=100$        & 0.151   & 0.086  & 0.043   &  0.028   &  0.124  & 0.440   & 0.332      & 0.166        &  0.152      &  0.371        \\
$T=250$        & 0.287   & 0.227  & 0.133   &  0.025   & 0.211    & 0.793   & 0.804      & 0.673        &  0.383      & 0.724         \\
$T=500$        & 0.457   & 0.431  & 0.327   &  0.032   & 0.341    & 0.961   & 0.983      & 0.967        &  0.733      & 0.947          \\
                   &           &          &            &            &              &           &              &                &                &                 \\
    \hline 
   \end{tabular}
\end{center}
\end{landscape}

\newpage
\fontsize{10pt}{10pt}\selectfont
\begin{figure}
\begin{center}
\caption{Sample path for DGP1-1}
   \includegraphics[width=11cm,height=6cm,clip]{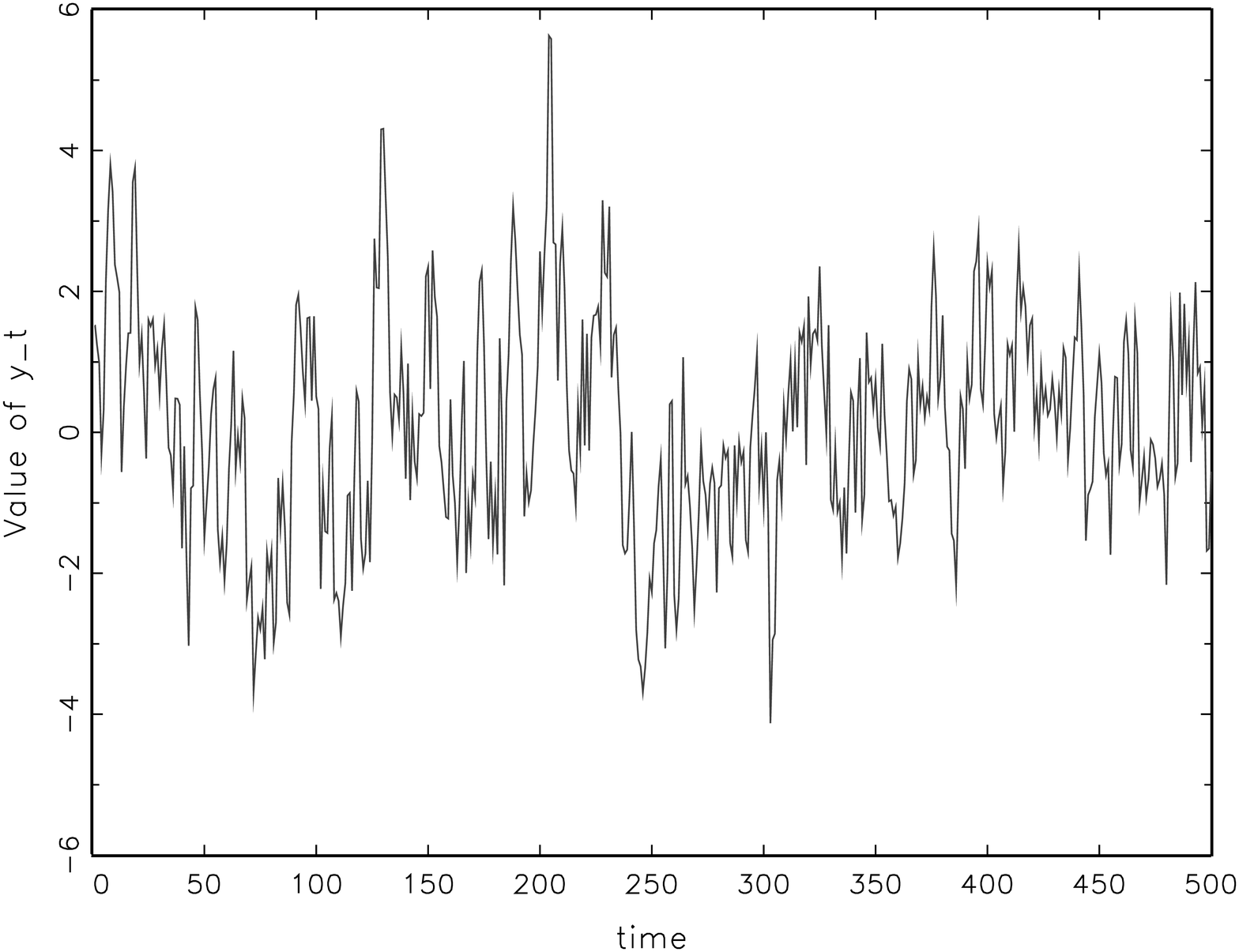} \\
\caption{Sample path for DGP1-1 with $\gamma_1=0.3$}
   \includegraphics[width=11cm,height=6cm,clip]{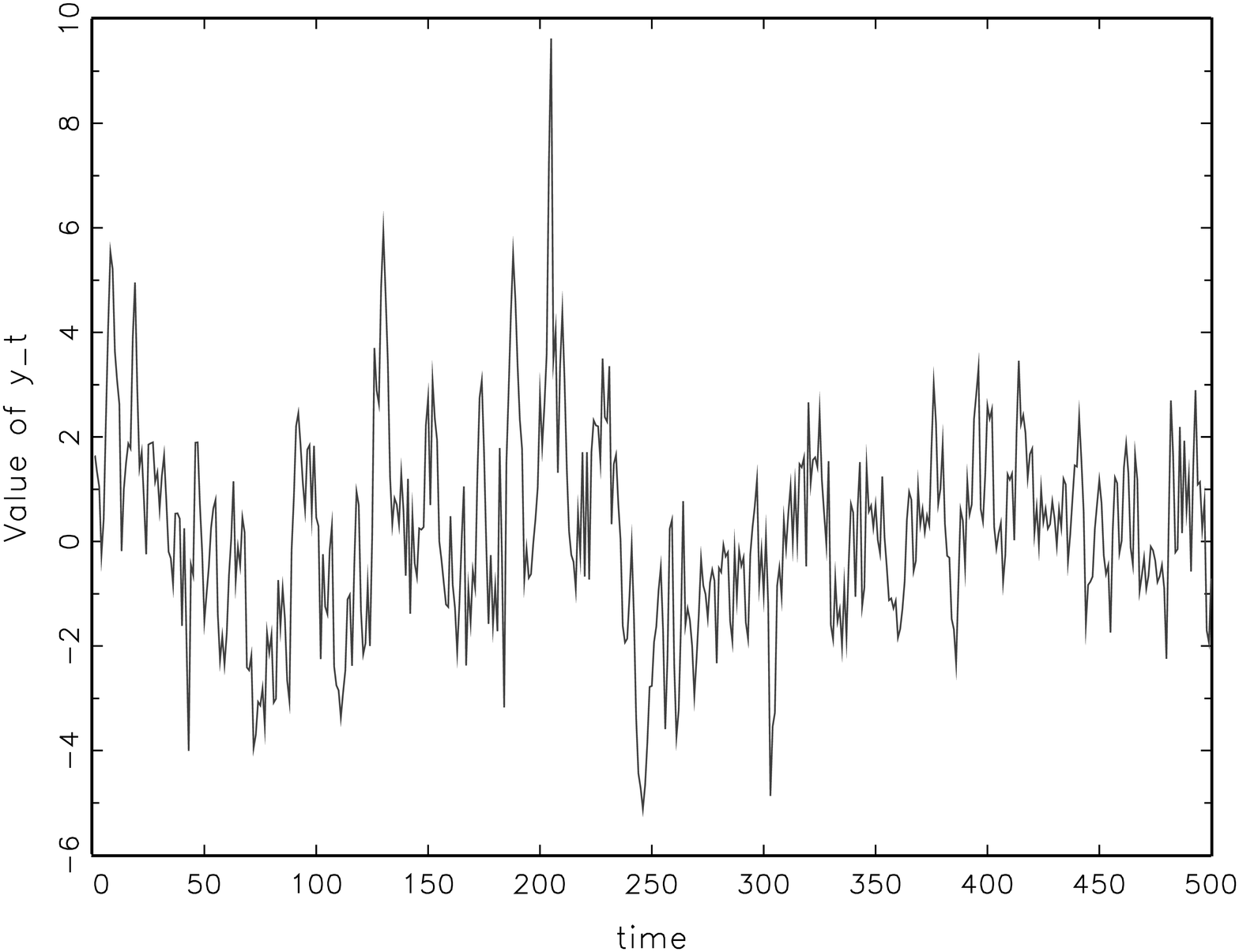} \\
\caption{Sample path for DGP2-3}
   \includegraphics[width=11cm,height=6cm,clip]{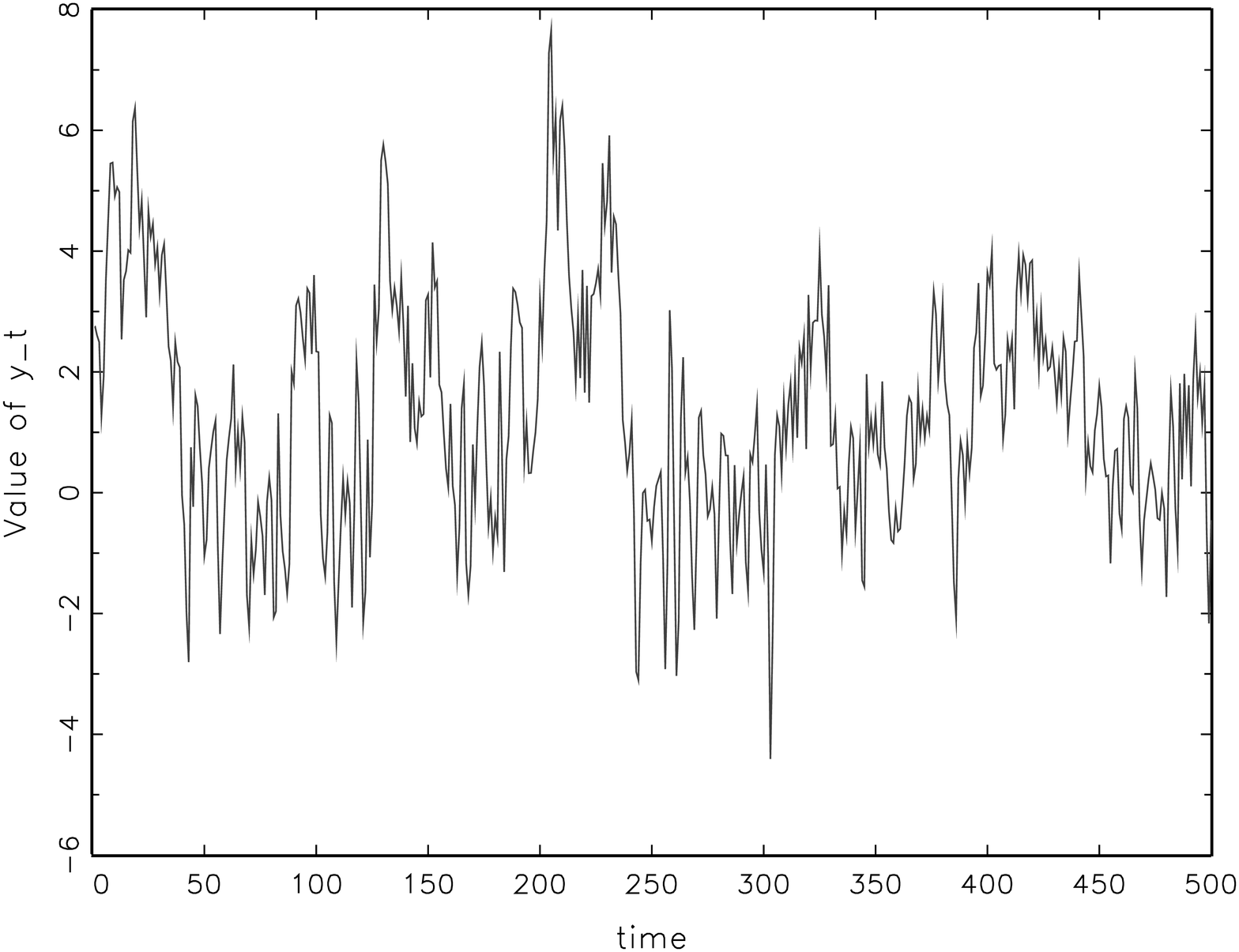} \\
\end{center}
\end{figure}

\end{document}